# Magnetic properties of poly(trimethylene terephthalate-*block*-poly(tetramethylene oxide) copolymer nanocomposites reinforced by graphene oxide-Fe$_3$O$_4$ hybrid nanoparticles

Anna Szymczyk[a], Sandra Paszkiewicz[b], Janusz Typek[a], Zdenko Špitalský[c], Izabela Janowska[d], Grzegorz Żołnierkiewicz[a], Nikos Guskos[a]

[a] West Pomeranian University of Technology, Institute of Physics, Piastów Av. 48, PL-70310 Szczecin, Poland

[b] West Pomeranian University of Technology, Institute of Material Science and Engineering, Piastów Av. 19, PL-70310 Szczecin, Poland

[c] Polymer Institute, Slovak Academy of Sciences, Dúbravská cesta 9, 845 41 Bratislava 45, Slovakia

[d] Institute of Chemistry and Processes for Energy, Environment and Health (ICPEES), CNRS and University of Strasbourg, 25 rue Becquerel, 67087 Strasbourg Cedex 2, France

*Corresponding author. Tel.: +48 91 449 45 81; e-mail address: aszymczyk@zut.edu.pl

**Abstract:** Thermoplastic elastomeric nanocomposites based on poly(trimethylene terephthalate-*block*-poly(tetramethylene oxide) copolymer (PTT-PTMO) and graphene oxide-Fe$_3$O$_4$ nanoparticle hybrid were prepared by *in situ* polymerization. Superparamagnetic GO-Fe$_3$O$_4$ hybrid nanoparticles before introducing to elastomeric matrix were characterized by X-ray photoelectron spectroscopy (XPS), X-ray diffraction (XRD), thermogravimetric analysis (TGA) and scanning electron microscopy (SEM). The effect of loading (0.3 and 0.5 wt %) of GO-Fe$_3$O$_4$ nanoparticle hybrid on the phase structure, tensile and magnetic properties of synthesized nanocomposites was investigated. The phase structure of nanocomposites was evaluated by differential scanning calorimetry (DSC) and dynamic mechanical thermal analysis (DMTA). Dispersion of GO-Fe$_3$O$_4$ nanoparticles in elastomeric matrix was evaluated by transmission electron microscopy (TEM). Magnetic properties of GO-Fe$_3$O$_4$ nanoparticle hybrid and nanocomposites with their content were characterized by using two different techniques: dc SQUID magnetization measurements as a function of temperature (from 2 to 300 K) and external magnetic field and ferromagnetic resonance (FMR) at microwave frequency.

## 1. Introduction

The interest in magnetic nanoparticles based on iron oxides is enormous due to their wide range of potential applications, especially in biomedicine, catalysis, waste water treatment, energy storage, and spintronics [1-9]. Magnetite ($Fe_3O_4$) and maghemite ($\gamma$-$Fe_2O_3$) are the most important ferromagnetic compounds among all iron oxides [10]. Saturation magnetization of magnetite reaches 92 emu g$^{-1}$ at room temperature (RT) in bulk material. On the other hand, maghemite is more stable than magnetite, more biocompatible, although presenting bulk saturation magnetization slightly smaller. The chemical formulae for stoichiometric magnetite can be written as $(8Fe^{3+})_{tetra}[8Fe^{2+}+8Fe^{3+}]_{octa}32O^{2-}$, where the subscripts tetra and octa mean tetrahedral (A sites) and octahedral (B sites) interstitials, respectively. Below 851 K magnetite is ferrimagnetic with A-site magnetic moments aligned antiparallel to the B-sites. At 120 K magnetite undergoes a first-order phase transition (Verwey transition), with a change of crystal structure, latent heat and decrease of the dc conductivity. The distribution of $Fe^{3+}$ and $Fe^{2+}$ in B sites changes from a dynamic disorder to a long-range order with an orthorhombic symmetry below 120 K. To study the magnetic properties of magnetite nanoparticles different experimental techniques were used. The basic and most often applied is the static (dc) and dynamic (ac) magnetometry employing vibrating sample and SQUID method. Two resonance techniques - ferromagnetic resonance (FMR) and Mössbauer spectroscopy usually deliver complementary magnetic characteristics [11-29].

Recent intense research on graphene and its derivatives follows the trend of exploration of novel low dimensional systems and is driven by possible promising applications of graphene-based magnets in spintronics. However, due to delocalized π bonding network graphene lacks localized magnetic moments and is intrinsically non-magnetic. Magnetic resonance spectroscopy measurements have confirmed theoretical predictions that in a single-layer graphene the carrier mediated exchange interaction leads to antiferromagnetic coupling [30]. The same experimental technique was used to study the magnetic correlations at zig-zag edges of a single-layer graphene [31]. It was found that the zig-zag spins are ferromagnetically correlated. Point defects such as the zig-zag edge states, vacancies with the trapped electron or chemical doping of foreign atoms, can introduce magnetic moments, but make the structure stability of graphene lower. Despite the efforts of increasing the magnetism of graphene materials, they usually display a weak Curie-type paramagnetism and magnetization below 0.2 emu/g. Recently, it has been demonstrated that N-doping of graphene oxide (GO) can dramatically increase its magnetization up to 1.66 emu/g and makes the magnetism of GO change from purely spin-half paramagnetism to ferromagnetism with Curie temperature of ca. 100 K [8]. In another work it has been shown that it is also possible to inject spins to a single layer graphene from an adjacent ferromagnetic material (dynamical spin injection) [32]. A pure spin current was pumped into graphene from the processing magnetization of the ferromagnet with very high efficiency.

Many researchers have shown that integrating GO or reduced graphene oxide (RGO) with inorganic nanoparticles allows for the properties of the nanocomposite to be engineered for specific applications. For example, a new type of hybrid material – graphene/$Fe_3O_4$ – has been recently synthesized and found a broad range of applications e.g. in targeted drug delivery, magnetic resonance imaging, lithium-ion batteries, ion removal, sensor, catalysts, etc. [32, 33]. These multifunctional nanomaterials combine the beneficial effects of graphene - high conductivity and large surface-to-volume ratio - with strong magnetism, low price and low environmental toxicity of magnetite. Magnetic properties of GO-$Fe_3O_4$ nanocomposites have been investigated by measuring magnetization of produced samples in an external magnetic field [34-37]. At room temperature (RT) no magnetic hysteresis loop was observed, as expected, because the nanocomposite's blocking temperature was below RT. Saturation magnetization smaller than in magnetite nanoparticles was explained by low loading of $Fe_3O_4$ in nanocomposite. In each studied nanocomposite monodispersed magnetite nanoparticles were successfully attached to graphene sheets [37].

Graphene magnetic nanocomposites include also more complex systems such as Cu/RGO/$Fe_3O_4$ [38, 39, 40], Ag/RGO/$Fe_3O_4$ [41], Pd/RGO/$Fe_3O_4$ [42, 43] hybrids which were recently synthesized using plant extracts as biological materials under mild conditions. These hybrid nanocomposites can be used as magnetically separable and reusable catalyst for cyanation of aldehydes to nitriles [38] or for the reduction of 4-nitrophenol and organic days [39-43].

The dispersion of the highly exfoliated GO nanoplates decorated by $Fe_3O_4$ nanocrystals in polymer matrix is the crucial factor that determine the quality and reliability of nanocomposites and their expected magnetic properties. Shen et al. [44] have shown lightweight, multifunctional polyetherimide/graphene-$Fe_3O_4$ composites foams for shielding of electromagnetic pollution.

Our recent studies have shown that introduction of small amount (up to 0.5 wt. %) of highly exfoliated GO [45], graphene nanoplatelets (GNP) or hybrid system being a mixture of GNP and single-walled carbon nanotubes (SWCNT) [46-48] into polyester elastomeric nanocomposites improved their mechanical properties and can provide them new functional properties as electrical and thermal conductivity.

In this work, a superparamagnetic GO-$Fe_3O_4$ hybrid nanoparticles were prepared and used as nanofiller for polyester thermoplastic elastomeric matrix. The magnetic properties of the polyester thermoplastic polymer nanocomposite reinforced by GO-$Fe_3O_4$ nanoparticles hybrid were characterized by using two different techniques: dc SQUID magnetization measurements as a function of temperature and external magnetic field, and ferromagnetic resonance (FMR) at microwave frequency. The results of measurements are correlated with morphological features of the studied samples.

2. Experimental

2.1 Materials

Ferric chloride hexahydrate ($FeCl_3·6H_2O$), ferrous chloride tetrahydrate ($FeCl_2·4H_2O$), ammonium hydroxide, dimethyl terephthalate (DMT), poly(tetramethylene oxide) glycol (PTMG) with molecular weight of 1000 g $mol^{-1}$, tetrabutyl orthotitanate (TBT) were delivered by Sigma-Aldrich and used as received. Bio-1,3-propanediol (bio-PDO, Susterra®Propanediol) was purchased from DuPont Tate&Lyle (USA). Antioxidant Irganox 1010 was purchased from Ciba-Geigy (Switzerland). The GO nanoplatelets used for preparation of nanoparticle hybrid were prepared by a modified Hummers method [49]. For GO synthesis the expanded graphite with an average size of 5 μm from (SGL Carbon, Germany) was used.

**2.2 Preparation of GO-$Fe_3O_4$ hybrid structures and their characterization**

The GO-$Fe_3O_4$ hybrid nanoparticles were synthesized by inverse chemical co-precipitation method according to the following procedure [50]. 50 mg of GO powder was exfoliated in 150 mL of deionized water by ultrasonic treatment for 30 min. and followed by the addition of 1.6 g $FeCl_3 \cdot 6H_2O$ and 0.6 g $FeCl_2 \cdot 4H_2O$. The mixture solution was added dropwise into 250 mL three-necked flask containing 25 mL of 30 % ammonium hydroxide under vigorous mechanical stirring. To prevent oxidative reaction, the system was kept under nitrogen atmosphere. The resulting black solution was maintained at 80 °C for 1 h and then cooled to RT. The black precipitate was isolated by a permanent magnet and washed three times with deionized water and three times with ethanol. The collected magnetic material was dried at 50 °C in vacuum.

**2.3 Synthesis of PTT-PTMO/GO-$Fe_3O_4$ nanocomposites**

PTT-PTMO/GO-$Fe_3O_4$ nanocomposites were synthesized by a two-step polycondensation reaction of dimethyl terephthalate (DMT, Sigma-Aldrich), bio-1,3-propanediol (bio-PDO) and poly(tetramethylene oxide) glycol (PTMG) in the presence of TBT as catalyst and Irganox 1010 as an antioxidant. Details of the preparation method of PTT-PTMO based nanocomposites could be found in our previous publications concerning GO [45] and a hybrid system of GNP and SWCNT [47]. Before carrying the polymerization, a dispersion of GO-$Fe_3O_4$ has been prepared by dispersing the appropriate amount of functionalized graphene nanoplatelets in bio-PDO through ultrasonication for 15 min using laboratory homogenizer (Sonoplus HD 2200, with frequency of 20 kHz and 75% of power 200W) and subsequent intensive mixing for 15 min using high-speed stirrer (Ultra-Turax T25).

The content of rigid PTT and soft PTMO segments was approximately the same (i.e. 50 wt % of each).

**2.3 Characterization**

The synthesized GO-Fe$_3$O$_4$ has been characterized by X-ray photoelectron spectroscopy (XPS) performed on Kratos Axis ULTRA X-Ray photoelectron spectrometer equipped with monochromatic Al K$\alpha$ (h$\nu$=1486.6 eV) radiation to quantitatively analyze the chemical composition of the nanofiller.

X-ray diffraction (XRD) data were obtained using the X'Pert PAN analytical diffractometer employing Cu K$_\alpha$ radiation ($\lambda$=0.15406 nm) at an operating voltage and current of 40 kV and 40 mA, respectively. The nanocrystal size (*d*) was estimated according to Sherrer's equation [51]

$$d = K\lambda/B cos\theta \qquad (1)$$

where *K* is the Scherrer constant (0.89), *B* is the full width at half maximum (FWHM) of the strongest reflection peak, and *θ* is the diffraction angle.

Thermogravimetric analysis (TGA) was performed using a SETARAM TGA 92-16 instrument in air flow (20 mL min$^{-1}$) using a heating rate of 10 K min$^{-1}$ in a temperature range from 298 K up to 1073 K. Indium and aluminum were used for temperature calibration. The amount of applied samples was ~ 10 mg. Two paralleled runs were performed for each sample. The surface morphology of the GO and GO-Fe$_3$O$_4$ was observed by Dual Beam (FIB/SEM) Microscope Quanta 3D 200i (FEI).

The microstructure of nanocomposites was observed by transmission electron microscopy (TEM) on Topcon 002B - UHR microscope working with an accelerated voltage of 200 kV and a point-to-point resolution of 0.17 nm. Prior to the analysis thin slices of polymers were prepared by cryo-cutting technique and deposited on the copper grid.

The thermal transitions of neat copolymer and prepared nanocomposites were investigated with differential scanning calorimetry (DSC) on a Q100 thermal analyzer (TA Instruments). A standard heat-cool-heat program with heating/cooling rate of 10 K min$^{-1}$ was performed between 173 and 523 K. The cooling and second heating scans were used in order to determine the glass transition temperature, crystallization and melting peaks. The degree of crystallinity ($x_c$) has been calculated using the following equation:

$$x_c = 100 \; x \; (\Delta H_m / \Delta H_m^o) \qquad (2)$$

where $\Delta H_m$ is the melting peak area of an examined sample on DSC thermograms, and $\Delta H_m^o$ is the enthalpy of fusion (146 J/g [52]) of fully 100 % crystalline PTT.

The dynamic mechanical thermal analysis (DMTA) has been performed using a Polymer Laboratories MK II apparatus working in a bending mode. The samples were heated in the temperature range from 153 K to the polymer melting temperature and the heating rate of 3 K min$^{-1}$ at a frequency of 1 Hz was used.

The tensile and cyclic tensile properties of nanocomposites were performed on an Autograph AG-X plus (Shimadzu) tensile testing machine equipped with a 1 kN Shimadzu load cell, an contact optical long travel extensometer and the TRAPEZIUM X computer software, operated at a constant crosshead speed of 100 mm min$^{-1}$. Measurements on dumbbell samples were performed at RT with the grip distance of 20 mm according to DIN 53455 standard. The values of yield strength, stress and elongation at break, permanent set (with error bars representing the 95 % confidence interval) were calculated from the stress–strain curves.

MPMS-7 SQUID magnetometer was used for dc magnetization measurements in the 2 - 300 K temperature range in the zero-field-cooled (ZFC) and field-cooled (FC) modes and for isothermal ($T$ = 2, 300 K) magnetization in magnetic fields up to 70 kOe. Magnetic resonance study was carried out on a conventional X–band ($\nu$ = 9.4 GHz) Bruker E 500 spectrometer with the 100 kHz magnetic field modulation. The ferromagnetic resonance (FMR) spectra were taken in the 90 – 290 K range and were in the form of the first derivatives of microwave absorption with respect to the sweeping external magnetic field.

**3. Results and discussion**

**3.1 Structural characterization of GO-Fe$_3$O$_4$ nanoparticle hybrid**

XPS has been employed to analyze the nature and the relative amount of iron and oxygen-containing functional groups present on the graphene surface. C1s XPS spectrum of GO synthesized via Hummers method is as follows: $sp^2$-C: 6.09 %; $sp^3$-C: 42.26 %; C-O: 18.49 %; C=O: 24.32 %; O-C=O: 4.40 %; π-π: 1.16 %. For GO-$Fe_3O_4$, C1s XPS spectrum shows: $sp^2$-C: 56.59 %; $sp^3$-C: 23.91 %; C-O: 9.65 %; C=O: 4.57 %; O-C=O: 4.15 %; π-π: 1.13 %. Whereas, the O1s XPS spectrum (**Figure S1**) shows: $FeO_x$: 66.88 %, C=O/Fe-OH: 16.27 %; C-O: 13.27 %. After successful functionalization of $Fe_3O_4$, clearly visible changes in C1s XPS spectrum were observed. An increase of $sp^2$ hybridized carbons from 6.09 to 56.59 %, and at the very same time a decrease in $sp^3$ hybridized carbons from 42.26 to 23.91 %, can be noted. The XPS spectrum of GO-$Fe_3O_4$ consists of functional groups such as $sp^2$ (C=C), epoxy/hydroxyl (C-O), carbonyl (C=O) and carboxylates (O-C=O) [53]. Moreover, in the case of GO-$Fe_3O_4$, XPS spectra indicate the presence of $FeO_x$ in the amount of 66.88 %, which certainly points to the chemical incorporation of iron oxide.

In order to confirm the attachment of $Fe_3O_4$ nanocrystals on the surface of GO the thermogravimetric analysis in air was performed on as-received GO, GO-$Fe_3O_4$ hybrid and $Fe_3O_4$ nanoparticles. The $Fe_3O_4$ nanocrystals content in the GO-$Fe_3O_4$ hybrid can be estimated from the residual weight percentages. The total weight loss of the $Fe_3O_4$ nanoparticles is 3.3 % (**Figure 1a**) for the whole temperature range because of the removal of adsorbed physical and chemical water. As can be seen in Fig. 1a, the TGA curves of GO show a three-step weight loss between 273 to 1073 K with a total weight loss of ~98 %, while GO-$Fe_3O_4$ hybrid in this temperature range undergoes a weight loss of 29.6 %, which is closely related to the oxidation of GO and confirms high concentration of $Fe_3O_4$ nanocrystals (~70 wt %) on GO nanosheets.

**Figure 1b** shows X-ray diffractograms of GO-$Fe_3O_4$ and $Fe_3O_4$ nanoparticles synthesized in the same conditions. The diffraction peaks at 30.26, 35.60, 43.35, 53.76, 57.25, 62.95 ° correspond to the (220), (311), (400), (422), (511), (440) and are in good accordance with the inverse cubic spinel phase of $Fe_3O_4$ (magnetite, JCPDS card no. 85-1436) [51]. Using the most intense reflection (311), the average size of $Fe_3O_4$ nanocrystals at GO surface was calculated to be ca. 8.9 nm. In our earlier work it was shown that the GO powder has characteristic diffraction peak at 2θ ~ 10 ° attributed to the (001) reflexion [54]. The disappearance of this peak in XRD pattern of GO-$Fe_3O_4$ hybrid suggests that the layer stacking of the GO sheets has been destroyed by the loading of $Fe_3O_4$ nanocrystals.

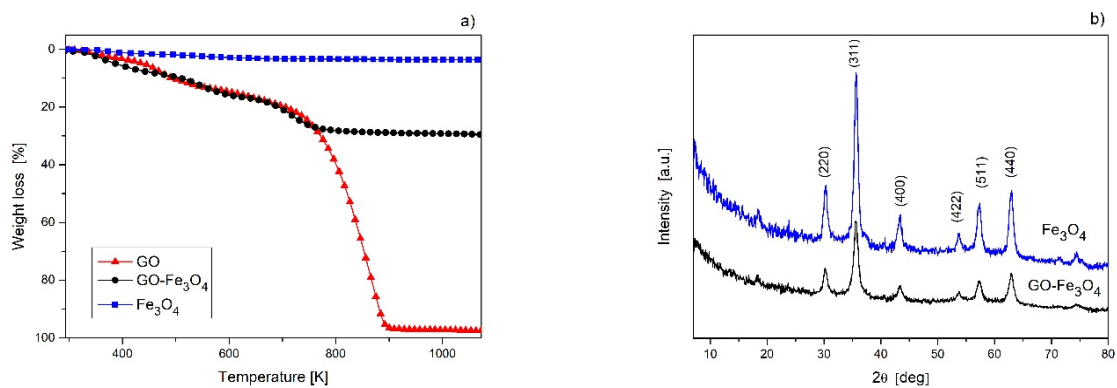

**Figure 1.** a) TGA curves of GO, GO-$Fe_3O_4$ nanoparticle hybrid and nano-$Fe_3O_4$; and b) X-ray diffraction pattern of GO-$Fe_3O_4$ and $Fe_3O_4$ synthesized in the same conditions.

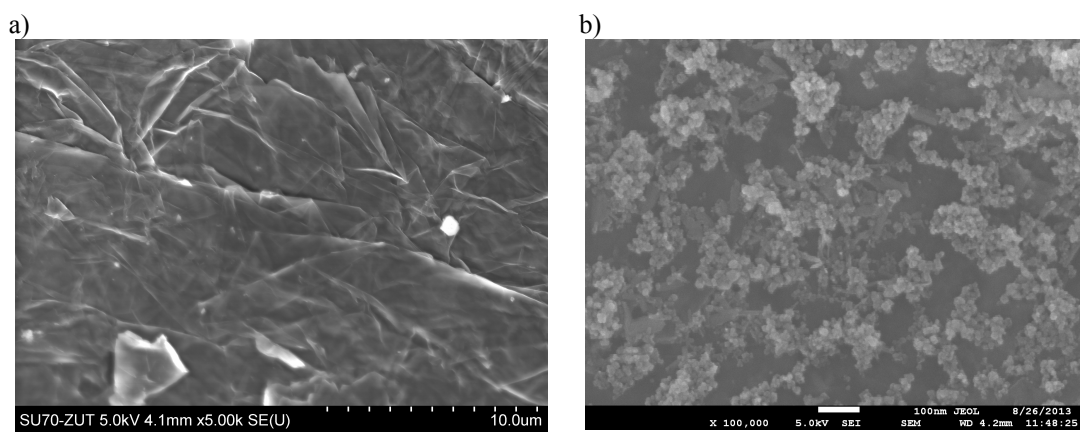

**Figure 2.** SEM images of as received GO (a) and GO-$Fe_3O_4$ (b).

The morphology of synthesized GO-$Fe_3O_4$ nanoparticle hybrid and GO used for their preparation was characterized by SEM. As shown in **Figure 2a**, the surface morphology of GO resembles strongly folded curtain, what indicates that GO flakes were rather overlapped than aggregated. On the other hand, when the synthesis of GO-$Fe_3O_4$ was carried out by an inverse chemical co-precipitation method, the formation of $Fe_3O_4$ took place on the graphene oxide sheets. SEM image (**Figure 2b**) shows that these $Fe_3O_4$ nanoparticles are densely distributed on GO sheets and they are not separated but decorated in bunches on the curled and thin wrinkled sheets, having a rather uniform distribution on the surface of GO. It can be observed from the SEM images that the average diameter of $Fe_3O_4$ nanocrystals at GO surface was around 10 nm, which is close to the value determined form XRD pattern.

### 3.2 Dispersion of GO-$Fe_3O_4$ in the PTT-PTMO matrix

The distribution of GO-$Fe_3O_4$ in PTT-PTMO copolymer matrix was studied by TEM microscopy. Representative images of nanocomposites containing 0.5 wt % of GO-$Fe_3O_4$ are presented in **Figure 3**. Nanocomposites exhibit rather uniform distribution of GO-$Fe_3O_4$ in the matrix (**Figures 3a-e**) and there were no large areas of graphene oxide that were not decorated with $Fe_3O_4$ nanoparticles. The high transparency of the GO sheets indicate that GO was well exfoliated into few layered sheets decorated by $Fe_3O_4$ nanocrystals. The observed morphology of GO-$Fe_3O_4$ at higher magnification have shown the sphere-like nanoparticles with sizes from 4 to 16 nm (**Figures 3b-c, e**) distributed on the surface of the GO sheets. Based on a total number of (N=124) particle sizes determined from several TEM images (**Figure 3f**) an average $Fe_3O_4$ particle size of 9 nm (2.1 nm) was estimated, which is in good agreement with the particles size determined form XRD data. Locally the areas with a more densely distributed nanoparticles (**Figure 3d**) were also observed. Likewise, the edges of GO flakes of several sheets can be clearly observed in **Figure 3c**. It can also be noted that $Fe_3O_4$ (**Figure 3**) were firmly attached to the graphene oxide sheets, that even a specimen's preparation for TEM analysis didn't affect the samples, which indicates that an excellent adhesion between GO and $Fe_3O_4$ particles has been achieved. Moreover, the presence of much smaller GO sheets densely decorated by $Fe_3O_4$ nanocrystals (**Figure 3d**) compared to the size of GO sheets used for preparation of these hybrid, was observed. Probably, due to high concentration of $Fe_3O_4$ nanocrystals (confirmed by TGA analysis) on GO sheets after inverse co-precipitation, some of them were cut into smaller pieces.

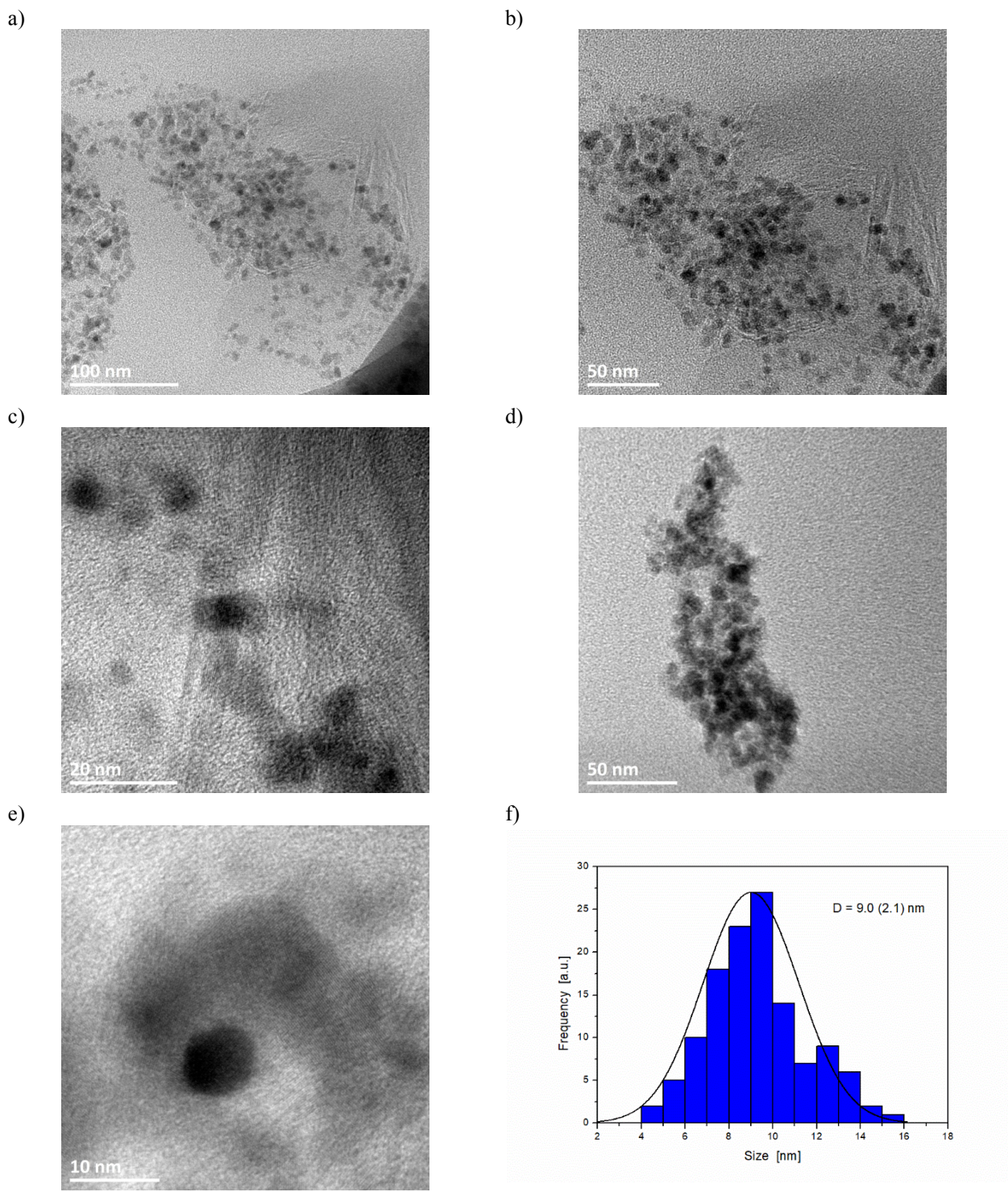

**Figure 3.** TEM images of the PTT-PTMO nanocomposites with 0.5 wt % of GO-$Fe_3O_4$ (a-e) and the histogram of distribution of the $Fe_3O_4$ particle sizes (f).

## 3.3 Effect of the presence of GO-$Fe_3O_4$ on phase structure and tensile properties of nanocomposites

In polyester block copolymers the heterophase structure, which can be reproducible in heating-cooling cycles, is responsible for their elastic and thermoplastic properties. The presence of GO-

$Fe_3O_4$ can affect their micro- and nanophase separation. When a synergy between the crystalline lamellae of polymer and the hybrid nanofiller is created, the improvement of the macroscopic properties of these composites can be expected. Values of intrinsic viscosity [$\eta$], $M_n$ and polydispersity (**Table 1**) of obtained composites are comparable to the values for the neat block copolymer. These results can indicate that at low loading of nanofillers *in situ* polymerization allows obtaining materials with comparable values of molecular weight, which have no effect on phase separation structure.

**Table 1.** Characteristics of the synthesized PTT-PTMO based nanocomposites

| Sample | $GO-Fe_3O_4$ [wt %] | [$\eta$] [dl g$^{-1}$] | $M_n$ [g mol$^{-1}$] | $M_w$ [g mol$^{-1}$] | $M_w/M_n$ |
|---|---|---|---|---|---|
| PTT-PTMO | 0 | 1.422 | 41 720 | 83 850 | 2.01 |
| PTT-PTMO/0.3GO-$Fe_3O_4$ | 0.3 | 1.423 | 42 440 | 87 850 | 2.07 |
| PTT-PTMO/0.5GO-$Fe_3O_4$ | 0.5 | 1.420 | 41 980 | 88 570 | 2.11 |

[$\eta$] - intrinsic viscosity, $M_n$ - number average molecular weight of the samples after filtration of nanofiller determined by SEC according procedure [44], $M_w/M_n$ – polydispersity

The effect of the presence of GO-$Fe_3O_4$ in PTT-PTMO matrix on the phase transitions (glass transitions, physical crosslink melting) in the obtained nanocomposites was studied by DMTA and DSC analysis. The storage modulus (E'), loss modulus (E") and loss factor (tan δ) as a function of temperature are presented in **Figure 4** for the neat block copolymer and its nanocomposites with GO-$Fe_3O_4$. Nanocomposites with GO-$Fe_3O_4$ show similar E' and E" profiles like the neat block copolymer matrices. Taking into consideration the experiment precision it can be inferred that the rubber moduli for nanocomposites are practically the same as for the neat block copolymer. The tan δ curves for all samples show two β - relaxations peaks, $β_1$ and $β_2$. These two relaxations are attributed to the glass transition of polyether-rich ($β_1$) phase and amorphous polyester ($β_2$) phase, respectively. It can be seen (**Figure 4, Table 2**), that the $β_1$- relaxation in nanocomposites is shifted to lower temperatures in comparison to the neat block copolymer. This confirms that addition of GO-$Fe_3O_4$ to the copolymer matrix improves the phase separation in the system. The final sample softening to a polymer melt, during which physical crosslinks (crystalline PTT domains) disrupt, takes place in the range of 450-475 K in all samples.

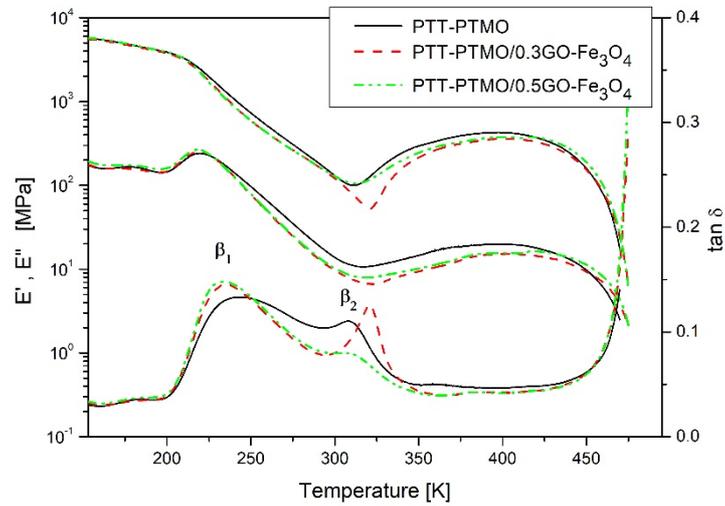

**Figure 4.** Storage and loss modulus and tan δ for neat PTT-PTMO copolymer and PTT-PTMO/GO-$Fe_3O_4$ nanocomposites containing of 0.3 and 0.5 wt % of GO-$Fe_3O_4$.

**Table 2.** DMTA and DSC data for the obtained nanocomposites

| Sample | $T_{\beta 1}$ [K] | $T_{\beta 2}$ [K] | $T_{g1}$ [K] | $T_m$ [K] | $\Delta H_m$ [J g$^{-1}$] | $T_c$ [K] | $\Delta H_c$ [J g$^{-1}$] | $\Delta T$ [K] | $x_c$ [%] |
|---|---|---|---|---|---|---|---|---|---|
| PTT-PTMO | 244 | 308 | 207 | 477 | 31.9 | 399 | 32.1 | 78 | 21.8 |
| PTT-PTMO/0.3GO-$Fe_3O_4$ | 235 | 321 | 205 | 477 | 31.1 | 402 | 31.4 | 75 | 21.3 |
| PTT-PTMO/0.5GO-$Fe_3O_4$ | 234 | 309 | 205 | 478 | 31.6 | 418 | 31.6 | 60 | 21.6 |

$T_{\beta 1}$, $T_{\beta 2}$ – temperatures of $\beta_1$ and $\beta_2$-relaxation corresponding to the glass transition temperatures determined from maximum of tan δ; $T_{g1}$ – glass transition temperature of amorphous soft phase; $T_m$, $\Delta H_m$ – temperature and enthalpy of melting of polyester phase; $x_c$ – degree of crystallinity; $T_c$, $\Delta H_c$ – temperature and enthalpy of crystallization of polyester phase, $\Delta T$ – degree of supercooling.

In **Figure 5** and **Table 2** the DSC results obtained during cooling and heating of the nanocomposites are presented. DSC data confirm conclusions obtained by DMTA. Herein, the addition of GO-$Fe_3O_4$ slightly shifted the value of $T_{g1}$ ($T_{\beta 1}$ on DMTA plots) toward lower temperatures. Moreover, no notable effect on the melting temperature ($T_m$) can be observed. In contrast, analysis of the cooling scans of nanocomposites and the neat block copolymer indicate that GO-$Fe_3O_4$ in PTT-PTMO copolymer could improve the crystallization of polymer matrix, since the crystallization temperature increased by 3-19 K. The degree of super cooling ($\Delta T=T_m-T_c$) of nanocomposites is smaller than that of neat PTT-TMO copolymer, which indicate that the presence of GO-$Fe_3O_4$ in PTT-PTMO block copolymer increases the rate of crystallization of PTT segments in these copolymers. This might be relevant to the processing properties of these materials. On the other hand, however, while the presence of nanoparticles accelerates the formation of crystalline phase, it does not correspond to the melting enthalpy, $\Delta H_m$, and designated on its basis the degree of crystallinity (values of $x_c$ are comparable to one another). It proves that GO-$Fe_3O_4$, randomly distributed in the copolymer matrix, promotes nucleation of PTT crystallites.

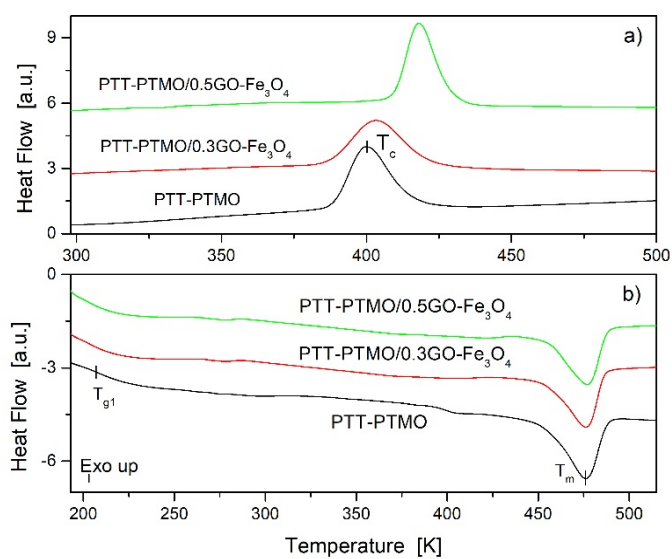

**Figure 5.** DSC thermograms obtained during cooling (a) and 2nd heating (b) for PTT-PTMO/GO-$Fe_3O_4$ nanocomposites.

As a result of the presence of GO-$Fe_3O_4$ and enhancement of phase separation in PTT-PTMO matrix at their low loading, the improvement of tensile and cyclic tensile properties of nanocomposites was observed. The representative stress-strain curves for nanocomposites obtained in the cyclic and non-cyclic uniaxial tensile tests are presented in **Figure 6** and the results are summarized in **Table 3**. The yield stress and tensile strength as well as yield strain and stress at break of nanocomposites slightly increases with the GO-$Fe_3O_4$ loading in the copolymer matrix. The values of permanent set (PS) in tension direction resultant from the attained strain of 100 and 200 % (**Table 3, Figure 6b**) for nanocomposites are close (within errors range) to the values obtained of the neat block copolymer.

**Table 3.** Tensile properties of the PTT-PTMO/GO-$Fe_3O_4$ nanocomposites

| Sample | $\sigma_y$ (MPa) | $\varepsilon_y$ (%) | $\sigma_b$ (MPa) | $\varepsilon_b$ (%) | PS(100) (%) | PS(200) (%) |
|---|---|---|---|---|---|---|
| PTT-PTMO | 12.1 ±0.1 | 41.7 ±0.1 | 23.5 ±0.2 | 622 ±12 | 38 ±2 | 95 ±3 |
| PTT-PTMO/0.3GO-$Fe_3O_4$ | 12.4 ±0.1 | 42.7 ±0.1 | 24.8 ±0.6 | 668 ±24 | 38 ±1 | 93 ±2 |
| PTT-PTMO/0.5GO-$Fe_3O_4$ | 12.6 ±0.2 | 43.2 ±0.3 | 26.8 ±0.9 | 680 ±31 | 38 ±2 | 95 ±3 |

$\sigma_y$, $\varepsilon_y$ – yield stress and strain respectively; $\sigma_b$ - stress at break; $\varepsilon_b$ -strain at break. PS(100), PS(200) – permanent set was taken as the strain at which zero load was measured on the first unloading cycle to 100% and 200% strain, respectively.

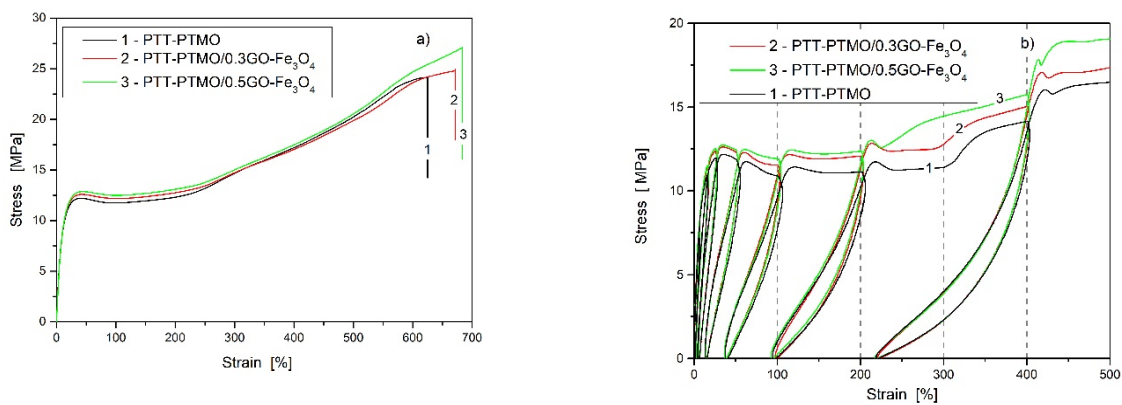

**Figure 6**. Stress-strain curves for PTT-PTMO/GO-$Fe_3O_4$ nanocomposites.

### 3.4 Magnetic properties of GO-$Fe_3O_4$ nanocomposites

The magnetic properties of PTT-PTMO/GO-$Fe_3O_4$ nanocomposites and GO-$Fe_3O_4$ nanoparticles used for their preparation were studied using a SQUID magnetometer in order to investigate the temperature dependence of dc magnetic susceptibility and isothermal magnetization as a function of magnetic field as well as a magnetic resonance spectrometer to examine the ferromagnetic/paramagnetic spectra at microwave frequency.

When sample is cooled to the lowest temperature without external magnetic field (ZFC mode), the magnetic moments of each nanoparticle align along the easy axis in the lattice, and since the crystallites are oriented randomly, the overall magnetic moment of sample will be zero. As the temperature is increased, thermal energy releases some magnetic moments from the easy axis and they will be align along the external field. The magnetocrystalline energy $K \cdot V$ (where $K$ is magnetocrystalline anisotropy energy density constant and $V$ the volume of a nanoparticle) plays a dominate role in that temperature range. At a specific temperature ($T_B$ - blocking temperature) the largest number of moments is aligned along the external field and gives maximum magnetization. Often a simple relation between magnetocrystalline and thermal energy is used to estimate $T_B$ [55]:

$$K \cdot V = 25 \, k \, T_B \qquad (3)$$

where $k$ is Boltzmann's constant. Above $T_B$ the Zeeman energy $\mu \cdot H$ (where $\mu$ is the magnetic moment and $H$ external magnetic field) is smaller than the thermal energy what causes randomization of the moments and the net magnetization will decrease with the increase in temperature (superparamagnetic phase). On the other hand, when sample is cooled from high temperature in the presence small external field (FC mode) the decrease in thermal energy causes orientation of moments along the direction of the field and subsequent increase in magnetization. The ZFC and FC curves coincide or show the same tendency as the temperature is decreased. The situation changes on further cooling below $T_B$: the Zeeman energy overcomes the thermal energy and causes the moments to orient partially along the applied field, what results in separation of ZFC and FC magnetization curves. At the lowest temperature the Zeeman energy causes the maximum orientation of moments in the field direction and the biggest magnetization.

In **Figures 7– 9** the temperature dependence of magnetization in ZFC and FC modes is shown for GO-Fe$_3$O$_4$ nanoparticles and PTT-PTMO/GO-Fe$_3$O$_4$ nanocomposites. Four panels in each figure present $M(T)$ curves taken in four different magnetic fields: $H$ = 10, 100, 1000, and 10000 Oe. These figures display the characteristic thermal irreversibility anticipated for an assembly of single domain magnetic particles due to the blocking-unblocking process of magnetic moments during variation of the thermal energy. In general, all samples display a rather broad size distribution of magnetic nanoparticles what is evidenced by the broadness of $\chi_{ZFC}$ curves and a large separation of the blocking temperature $T_B$ (maximum in ZFC mode) and the irreversibility temperature $T_{irr}$ (temperature of splitting of $\chi_{ZFC}$ and $\chi_{FC}$ curves). The latter is associated with the blocking of the biggest particles. Another evidence of a broad size distribution is no saturation of FC magnetization at low temperature. Comparison of **Figure 8** with **Figure 9** shows a very similar thermal behavior of magnetizations in both nanocomposites PTT-PTMO/0.3GO-$Fe_3O_4$ and PTT-PTMO/0.5GO-$Fe_3O_4$. On the other hand, comparison of magnetic susceptibilities of both nanocomposite samples with GO-$Fe_3O_4$ sample reveals a significant difference – nanocomposites seems to display two blocking temperatures, because two shallow maxima can be discerned in $\chi_{ZFC}$ curves, while only one maximum is visible in $\chi_{ZFC}$ curve of GO-$Fe_3O_4$ sample. Thus, the procedure of synthesis of polymer samples affects to some degree their magnetic properties producing bimodal-like distribution of the blocking temperatures. A simple explanation of this phenomenon involves the presence of a magnetite-polymer interface that produces the core-shell structure of magnetite nanoparticles in these samples. It is also apparent that there is no evidence of Verwey transition in magnetization curves in the neighborhood of 120 K. There is another interesting feature seen in the temperature dependence of magnetic susceptibility in both block copolymer samples that is absent in GO-$Fe_3O_4$. At low temperatures (T < 20 K) and in high magnetic fields (H > 1 kOe) it is manifested in the upturn in dc susceptibility of PTT-PTMO/0.3GO-$Fe_3O_4$ and PTT-PTMO/0.5GO-$Fe_3O_4$. Similar behavior was observed for composites of $Fe_3O_4$ nanoparticles with alginic acids [16]. It was interpreted in terms of the excitation of the spin waves with discrete spectrum. At higher temperatures, the magnetic energy levels are broadened and form a continuous excitation spectrum. In that case the temperature dependence of magnetization can be the same as in bulk materials. The proposed interpretation of magnetization upturn at low temperature assumes the presence of small nanoparticles because the involved quantum effects are magnified in restricted dimensions. Bimodal-like distribution of blocking temperatures in our polymer nanocomposite samples and the presence of magnetite-polymer interface suggest other possible explanation of magnetization upturn at low temperature - ordering of spins in the nanoparticle shells in the magnetic field of already ordered spins in the magnetite cores [56].

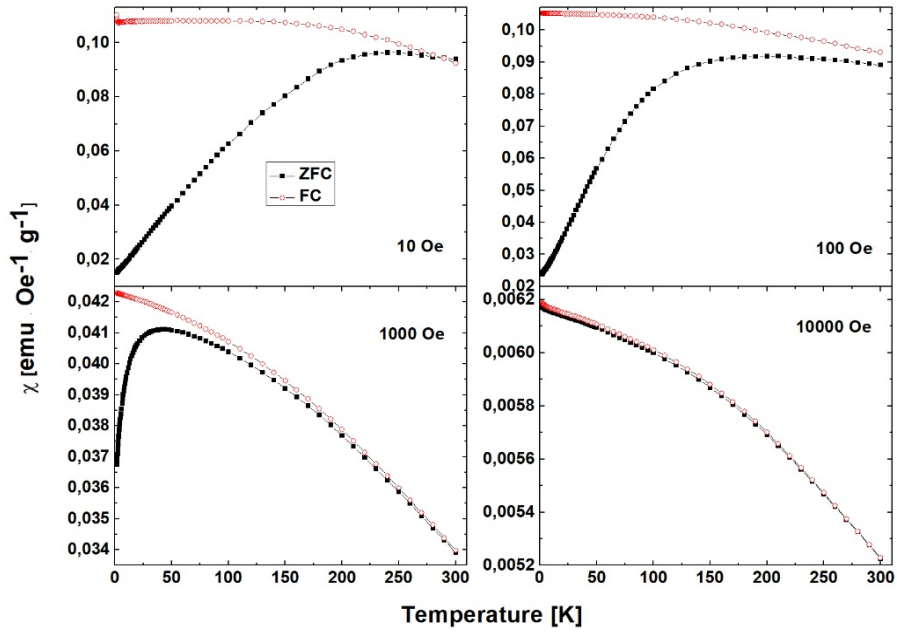

**Figure 7.** Temperature dependence of magnetic susceptibility in ZFC and FC modes in four magnetic fields (10, 100, 1000, and 10000 Oe) of GO-Fe$_3$O$_4$ sample.

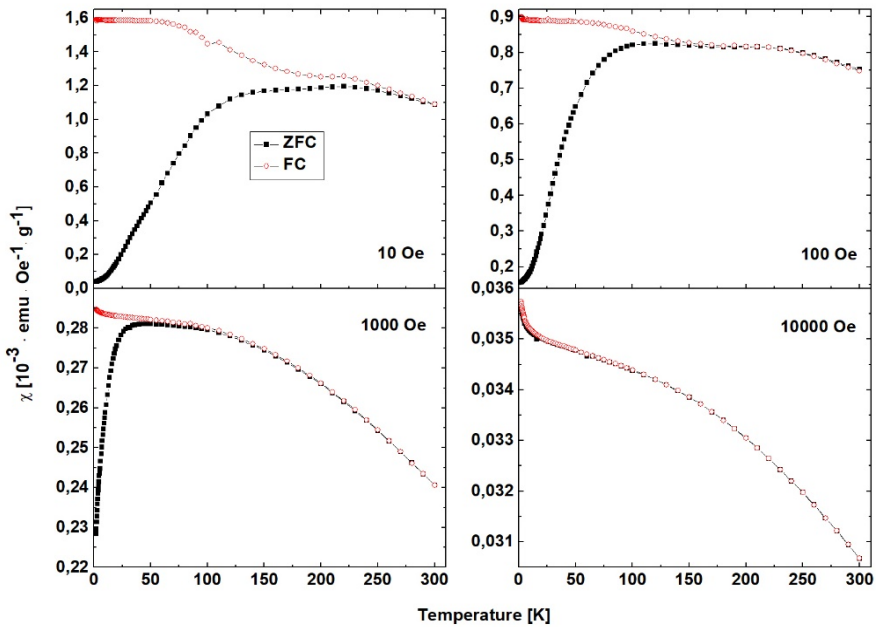

**Figure 8.** Temperature dependence of magnetic susceptibility in ZFC and FC modes in four magnetic fields (10, 100, 1000, and 10000 Oe) of PTT-PTMO/0.5GO-Fe$_3$O$_4$ sample.

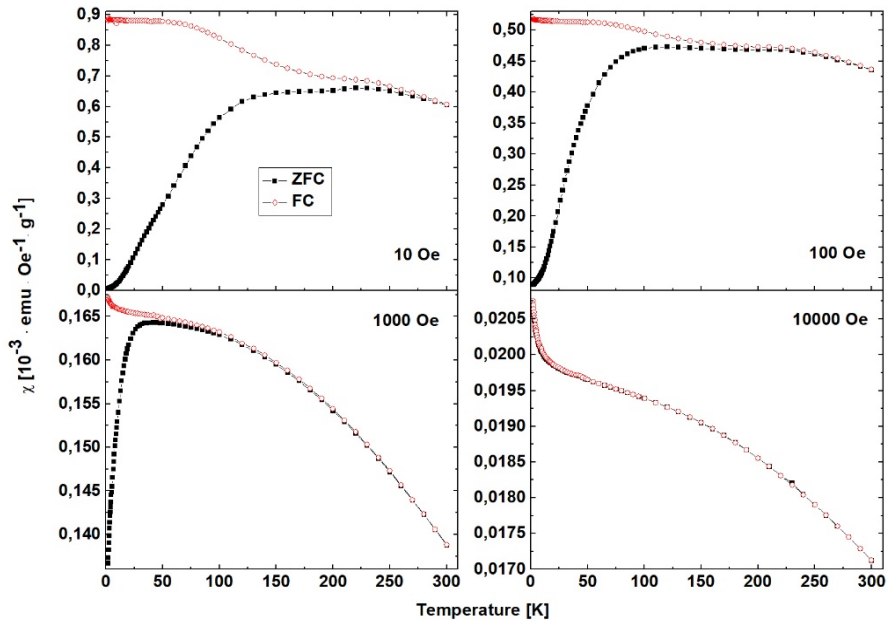

**Figure 9.** Temperature dependence of magnetic susceptibility in ZFC and FC modes in four magnetic fields (10, 100, 1000, and 10000 Oe) of PTT-PTMO/0.3GO-$Fe_3O_4$ sample.

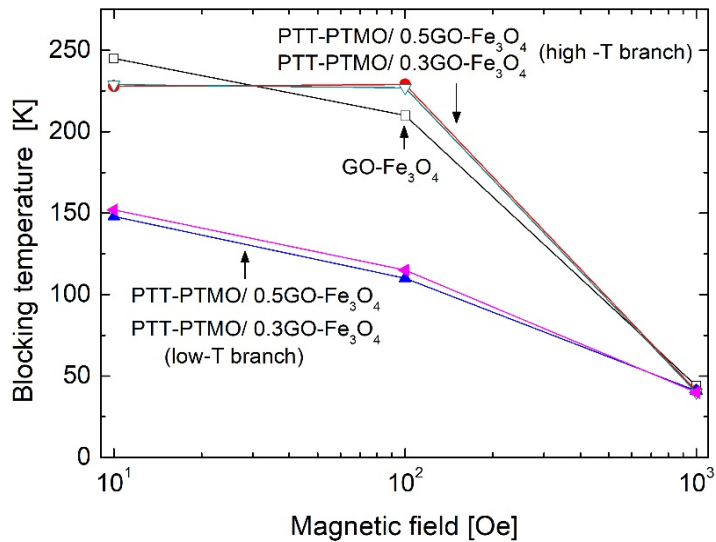

**Figure 10.** Dependence of the blocking temperature on external magnetic field in investigated GO-$Fe_3O_4$ and PTT-PTMO/GO-$Fe_3O_4$ nanocomposites. The solid lines are merely guides to the eyes.

The blocking temperature depends on applied magnetic field and is expected to shift to lower temperatures on increasing field strength because magnetic field lowers the barrier between two easy axis orientations. In **Figure 10** this dependence is presented for GO-$Fe_3O_4$ nanoparticles and nanocomposites with their content. For both nanocomposites, as there are two blocking temperatures, there will be two branches of $T_B(H)$ curves. It is easy to see that the high temperature branch for GO-$Fe_3O_4$ and PTT-PTMO nanocomposites coincides and only the low temperature branch is specific for both nanocomposites samples. Double peaked $M_{ZFC}$ could be explained to be the result of existence of two different subsystems of nanoparticles with different effective average sizes. The high temperature $T_B(H)$ branch represents one collection of larger nanoparticles in GO-$Fe_3O_4$ and PTT-PTMO nanocomposites and the low temperature branch is specific to PTT-PTMO nanocomposites. In the latter the core of a nanoparticle is effectively smaller because of the existence of a shell layer having different magnetic properties. Applying Eq. (3) to the common to all samples blocking temperature approximated to zero magnetic field ($T_B(0) = 250$ K) and using an average nanoparticle size obtained from TEM images ($D = 9$ nm), it can be calculated that $K = 2.2 \cdot 10^6$ erg cc$^{-1}$, what is much larger than that for bulk $Fe_3O_4$ ($1.35 \cdot 10^5$ erg cc$^{-1}$ at RT). This enhancement of the particle's anisotropy is likely associated with a surface having larger anisotropy (aggregation of nanoparticles).

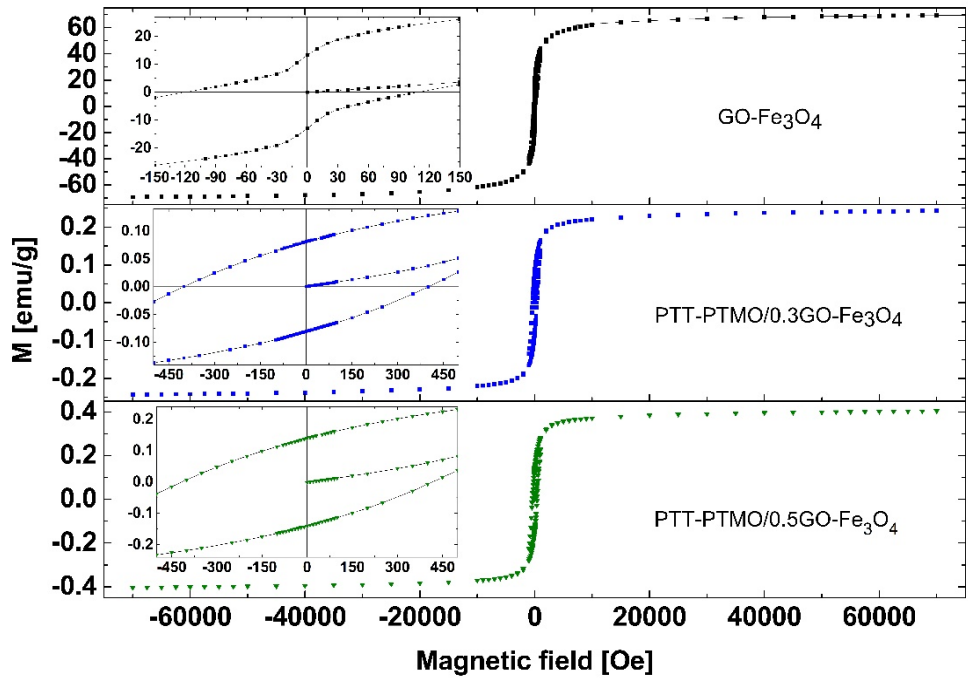

**Figure 11**. Isothermal magnetization at $T = 2$ K for GO-$Fe_3O_4$ and PTT-PTMO/GO-$Fe_3O_4$ nanocomposites. The solid line in the upper panel is the fit to the sum of ferromagnetic and paramagnetic components. The insets show expanded views of the low magnetic field behavior (the solid lines are guides for the eyes).

In **Figure 11** the isothermal (at $T = 2$ K) magnetizations of GO-$Fe_3O_4$ and PTT-PTMO/GO-$Fe_3O_4$ nanocomposites in magnetic fields up to 70 kOe are presented. As anticipated, magnetization in form of hysteresis loop is observed as the samples are in the blocked, ferromagnetic state. The diamagnetic contribution of polymer matrix has been subtracted in case of PTT-PTMO/0.3GO-$Fe_3O_4$ and PTT-PTMO/0.5GO-$Fe_3O_4$ samples. On the other hand, at RT no hysteresis loop is registered because this temperature is higher than the blocking temperature and the samples are in superparamagnetic state. The parameters of the observed loops (saturation magnetization, remanent magnetization, coercive field) at $T = 2$ K are listed in **Table 4.**

**Table 4.** Hysteresis loop parameters at $T = 2$ K of GO-$Fe_3O_4$ and PTT-PTMO/GO-$Fe_3O_4$ nanocomposites. Saturation and remanent magnetizations are expressed on a unit mass (1 g) of the whole sample, including GO and polymer masses.

| Sample | $H_c$ [Oe] | $M_R$ [emu g$^{-1}$] | $M_s$ [emu g$^{-1}$] | $M_R/M_s$ |
|---|---|---|---|---|
| GO-$Fe_3O_4$ | 114 | 13.2 | 69.3 | 0.19 |
| PTT-PTMO/0.3GO-$Fe_3O_4$ | 400 | 0.08 | 0.24 | 0.33 |
| PTT-PTMO/0.5GO-$Fe_3O_4$ | 415 | 0.14 | 0.40 | 0.35 |

The values of the saturation magnetization and the remanent magnetization in **Table 4** are calculated for a unit mass (1 g) of the whole sample, including GO and copolymer masses. Taking into account that in 1 g of GO-$Fe_3O_4$ sample the mass of magnetite is 0.686 g (this can be calculated from XPS data) it follows that the saturation magnetization would be 101 emu g$^{-1}$$_{(Fe3O4)}$. This is slightly more than for bulk magnetite at that low temperature. But this value has to be reduced because the registered magnetization of GO-$Fe_3O_4$ sample do not reaches saturation even in 70 kOe magnetic field. Thus, the observed signal were treated as composed from two components: ferromagnetic, saturated below



20 kOe and paramagnetic that will not be saturated in the highest applied magnetic fields. By subtracting the paramagnetic signal from the observed data, the remaining ferromagnetic magnetization could be obtained. Magnetization of the paramagnetic component was approximated by Langevin function. The classical Langevin expression describes magnetization of uncoupled array of magnetic moments or an exchange coupled superparamagnetic array of magnetic moments (when the direction of the easy axis does not play a role)

$$M = M_S \left( coth(x) - \frac{1}{x} \right), \qquad (4)$$

where $M_S$ is saturation magnetization, $x = \frac{\mu H}{kT}$ is the ratio of magnetic to thermal energy, $\mu$ is the magnetic moment and $k$ is Boltzmann's constant. Fitting of $M(H)$ curve for GO-Fe$_3$O$_4$ sample in 20 – 70 kOe range by the sum of Eq. (4) and a constant (representing already saturated ferromagnetic component) was very accurate and gave the saturation magnetization of the ferromagnetic component to be 99 emu g$^{-1}$$_{(Fe3O4)}$ and the paramagnetic moment of 2.7 $\mu_B$ (Bohr magnetons). The fit is shown in **Figure 11** (upper panel) by a solid line.

Comparison of the values of the other loop parameters ($H_c$ and $M_R/M_S$) for GO-Fe$_3$O$_4$ with both nanocomposite samples brings up a striking difference between them (**Table 4**). The coercive field is roughly three times and the $M_R/M_S$ ratio (the squareness ratio coefficient $Q$) is two times bigger in the PTT-PTMO/GO-Fe$_3$O$_4$ nanocomposites in comparison with GO-Fe$_3$O$_4$. For non-interacting, randomly oriented particles with uniaxial magnetocrystalline anisotropy $Q = 0.5$, while for particles with cubic magnetocrystalline anisotropy $Q = 0.831$ [57]. As the squareness ratio represents the fraction of blocked particles at a specific temperature, its value increase with decrease of temperature [58]. Small values of this parameters indicate that the individual particles are single domains and display a strong random anisotropy [59]. Reduction of $Q$ could be due to different reasons like inter-particle interaction, distribution of particle sizes, the presence of various defects, etc. Petrychuk et al. [60] investigated samples containing interacting magnetic nanoparticles forming clusters of different sizes. For small size magnetic clusters, the magnetic energy may be small enough to enable thermal energy the enhancement of the superparamagnetic contribution into magnetization. This results in small values of coefficient *Q*. On the other hand, for larger cluster the magnetic energy is high what leads to the corresponding enhancement of its superferromagnetic properties. The hysteresis loop becomes more rectangular and the squareness ratio coefficient increases. These considerations suggest that *Q* value might be considered as a measure of cluster sizes in a strongly interacting system of ferromagnetic nanoparticles. Enhancement of the coercive field in both nanocomposite samples in comparison to GO-Fe$_3$O$_4$ might be explained by bigger contribution of the shape anisotropy in the former samples [57, 61, 62]. It has been shown that in the case of magnetic nanowires the coercivity can be evaluated from expression containing two terms - one related to the shape anisotropy and the other to the magnetocrystalline anisotropy. In case of specifically arranged nanoparticle assemblies the first term can be dominating. This might be the case of our nanocomposite samples were chains of Fe$_3$O$_4$ nanoparticles would ensure high value of the coercive field. **Figure 12** shows magnetization curves of each sample registered at RT. The data is plotted as *M/M$_S$* (dimensionless) vs. *H* to compare the shapes of the curves.

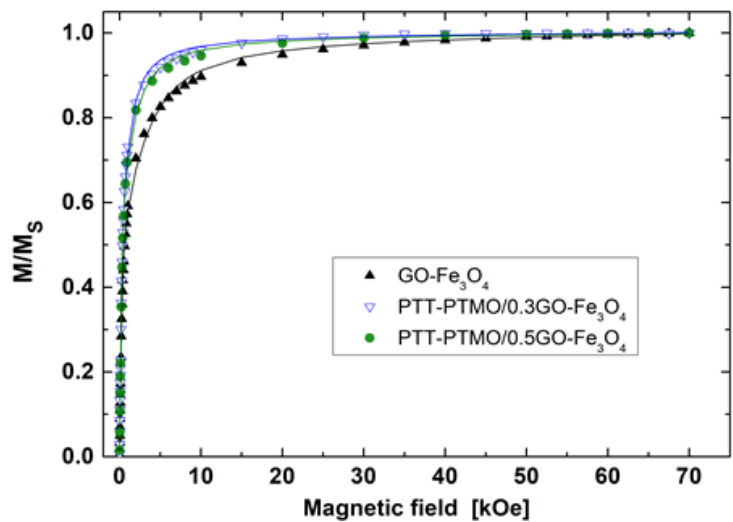

**Figure 12.** Normalized magnetization of GO-Fe$_3$O$_4$ and PTT-PTMO/GO-Fe$_3$O$_4$ nanocomposites in an external magnetic field at T = 300 K. The solid lines are the least-squares fits to the sum of two Langevin functions with different magnetic moments and a small term linear in magnetic field.

For PTT-PTMO/0.3GO-Fe$_3$O$_4$ and PTT-PTMO/0.5GO-Fe$_3$O$_4$ nanocomposites the curves are very similar, but GO-Fe$_3$O$_4$ nanoparticles hybrid showed a smaller magnetization in fields of intermediate strengths and a slower approach to saturation. The magnetization curves in **Figure 12** cannot be adequately fitted using a single Langevin function with a single magnetic moment. Satisfactory fits of the data (solid lines in **Fig. 12**) were obtained using the sum of two Langevin functions corresponding to two different magnetic moments (differing by a factor of 10) and an additional small term linear in magnetic field. The obtained magnetic moments probably do not have much physical meaning, there are only indication of a spread of magnetic moment values (and corresponding nanoparticle sizes) in our samples [63]. Reduction of magnetization in intermediate fields in case of sample GO-$Fe_3O_4$ might be explained by an inter-particle interaction causing formation of flux closer loops [64].

**Ferromagnetic resonance**. Selection of FMR spectra in the form of magnetic field dependence of the field derivative of the microwave absorption registered at different temperatures in 90 - 300 K range is presented in **Figure 13** for GO-$Fe_3O_4$ (upper panel) and PTT-PTMO/0.5GO-$Fe_3O_4$ nanocomposite (lower panel).

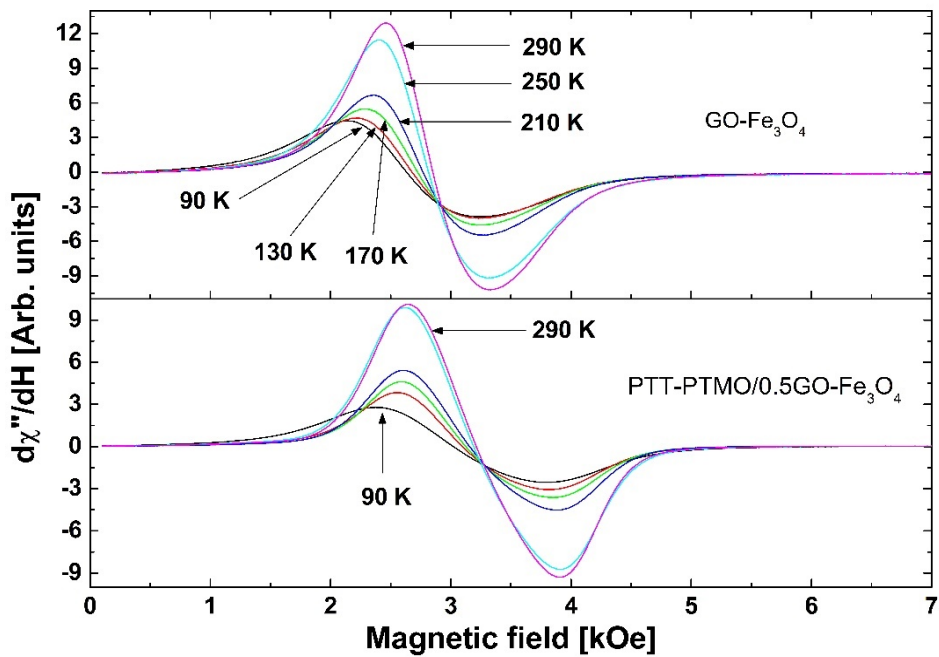

**Figure 13.** FMR spectra registered at different temperatures of GO-$Fe_3O_4$ (top panel) and PTT-PTMO/0.5GO-$Fe_3O_4$ nanocomposite (lower panel).

All FMR spectra are dominated by an intense, broad and asymmetrical line. The registered FMR spectra reflect very complex spatial arrangements of magnetic particles and a broad range of different interactions they are involved. To obtain information, albeit approximate and limited, about particles and their interactions, the spectra were decomposed on components described by Gaussian line shape functions [65, 66, 67]. Two Gaussian components were needed to obtain a very good agreement between the experimental and calculated spectra. As a result of fitting the values of the resonance field, linewidth and amplitude of each component were obtained at specific temperatures. Another important parameter – the integrated intensity – was calculated as the product of the amplitude and the square of linewidth. The results of fittings are presented in **Figures 14, 16,** and **17** which show the temperature dependence of the resonance fields, linewidths, and integrated intensities, respectively.

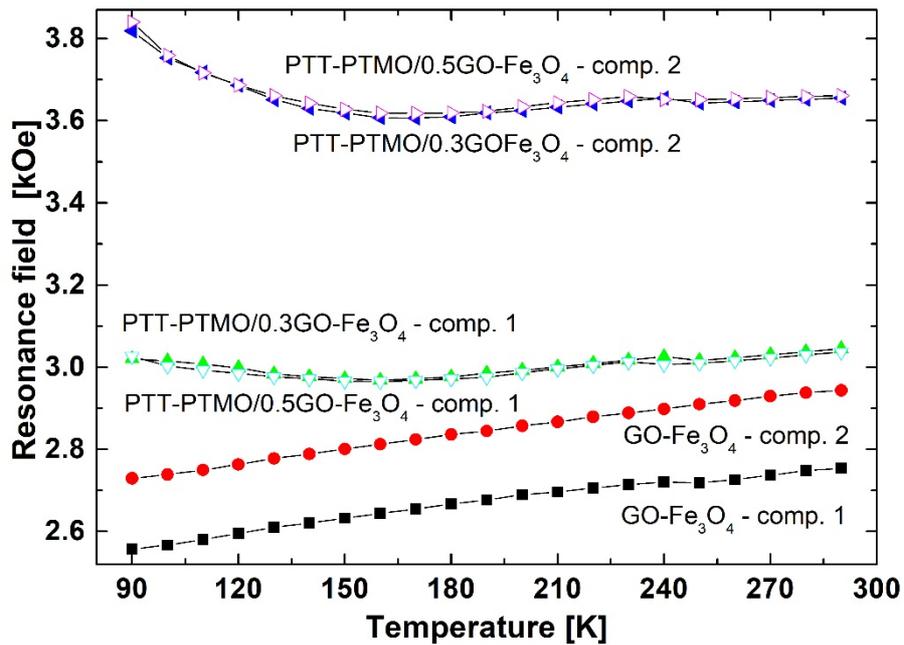

**Figure 14.** Temperature dependence of the resonance fields of the component lines (designated as comp. 1 and comp. 2) for GO-$Fe_3O_4$ and nanocomposites. The solid lines are guides for the eyes.

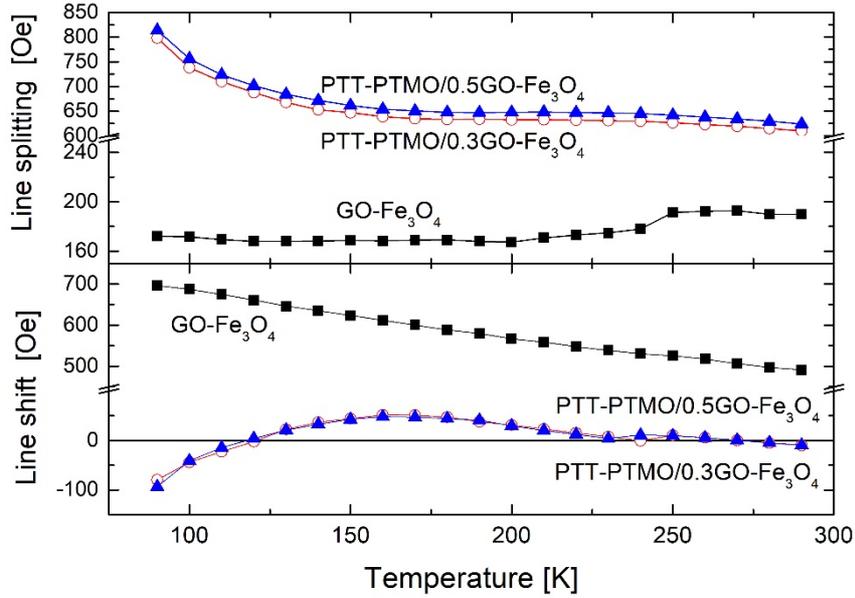

**Figure 15.** Temperature dependence of the line splitting between two components (top panel) and line shift (bottom panel) from magnetic field corresponding to $g = 2$ for GO-$Fe_3O_4$ and PTT-PTMO/ GO-$Fe_3O_4$ nanocomposites.

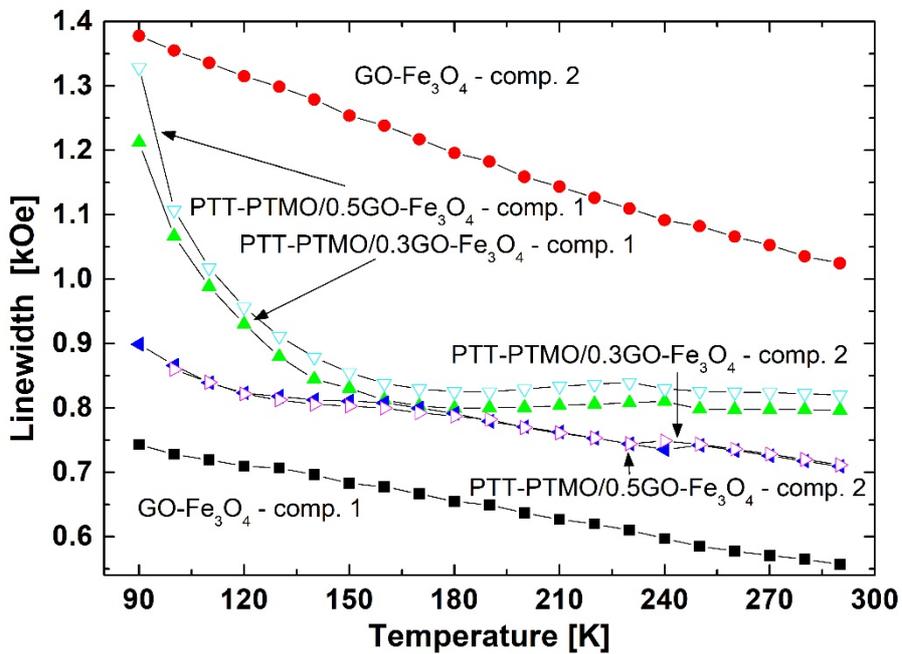

**Figure 16.** Temperature dependence of the linewidths of the component lines for GO-$Fe_3O_4$ and PTT-PTMO/GO-$Fe_3O_4$ nanocomposites. The solid lines are guides for the eyes.

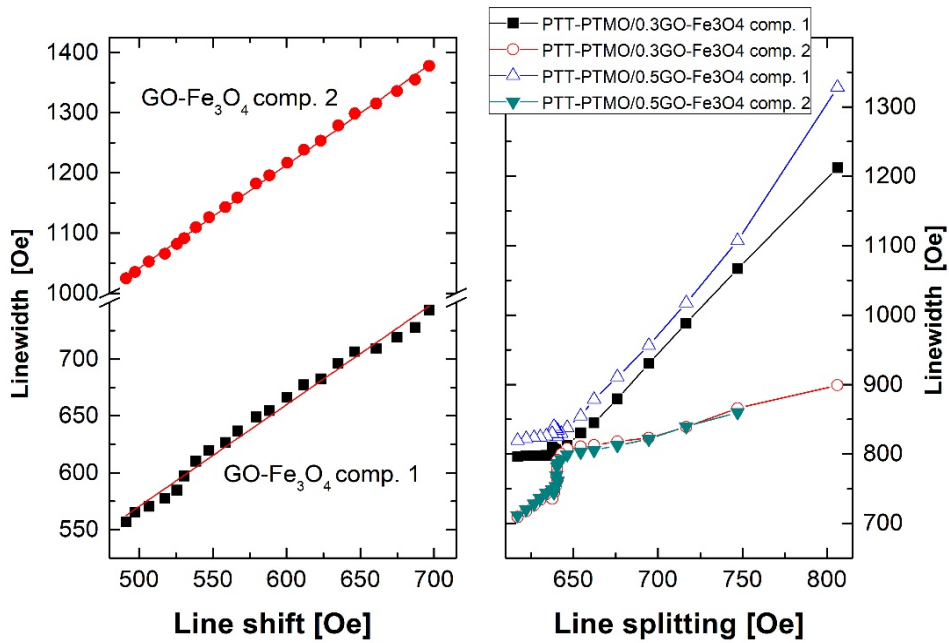

**Figure 17.** Dependence of component linewidth on the line shift from the reference field $H_{r0}$ in GO-$Fe_3O_4$ (left panel) and on the line splitting in PTT-PTMO/GO-$Fe_3O_4$ nanocomposites (right panel). Solid lines in the left panel are linear fits, in the right panel are guides for the eyes.

A remarkable difference between the two studied polymer nanocomposite samples on the one hand, and the GO-$Fe_3O_4$ on the other hand, is seen in **Figure 14** that illustrates the temperature dependence of the resonance field of each component (designated as comp. 1 and comp. 2) for three studied samples. Two characteristic features of these dependencies are easily to notice: there is a different splitting of the two components in polymer as compared to GO-$Fe_3O_4$ samples (large for the former and small for the latter) and there is a different shift of the split components from the magnetic field $H_{r0}$ corresponding to $g = 2$ (no internal field, superparamagnetic case) in both types of samples. These differences are illustrated in **Figure 15**. Top panel in **Fig.15** displays the temperature variation of the line split into two components in our three samples. Both polymer nanocomposites show a similar, large splitting of approximately 700 Oe, while that splitting in GO-$Fe_3O_4$ sample is significantly reduced and below 200 Oe. It is interesting to notice that similar differences were observed for the coercive fields in these samples what suggests a common mechanism behind those two different phenomena. Bottom panel in Fig. 15 shows the temperature dependence of the line shift defined as the difference between $H_{r0}$ (in our case $H_{r0}$ = 3340 Oe) and the magnetic field in the middle between comp. 1 and comp. 2 lines. Contrary to the case of line splitting this shift is large for GO-Fe$_3$O$_4$ (above 500 Oe and increasing with temperature decrease) and small (below 100 Oe) for both polymer nanocomposites. The reasons for these differences should be searched in morphological differences between GO-Fe$_3$O$_4$ and both PTT-PTMO/GO-Fe$_3$O$_4$ nanocomposites. In the following the large shift of the resonance line from $H_{r0}$ in GO-Fe$_3$O$_4$ will be explained by agglomeration of magnetite nanoparticles in that sample and the large split of two components in both polymer nanocomposites by a significant contribution of the surface energy due to a strong polymer and magnetite interface.

Study of temperature dependence of the resonance field in FMR experiments could provide valuable information on spin dynamics of ferromagnetic nanoparticles [68, 69, 70]. In FMR studies the following resonance condition must be satisfied: $\omega_0 = \gamma \cdot H_{eff}$, where $\omega_0$ is the microwave frequency, $\gamma$ the gyromagnetic ratio for electron, $H_{eff}$ the effective magnetic field which is the vector sum of externally applied field and an intrinsic filed. The latter field could have a few components, like magnetocrystalline anisotropy field $H_a$, shape anisotropy field $H_S$, dipolar field $H_D$, exchange field $H_{ex}$, etc. The number of field components depends on physical and chemical characteristics of particular nanomaterial. Considering FMR line splitting observed in our samples, an expression containing three terms is necessary: the Zeeman term, and two terms describing different anisotropies: surface anisotropy and bulk (volume) magnetocrystalline anisotropy. The volume term, proportional to $K_V$ (volume anisotropy constant) is standard and may be regarded as a uniform intrinsic field $H_a$ independent on particle size. There is a simple relation between these quantities

$$H_a = \frac{2K_V}{M} \qquad (5)$$

where $M$ is magnetization. The surface anisotropy term contains $K_s$ (surface anisotropy constant) and acts like an additional uniform field. Its strength depends on particle size the smaller the particle the stronger that field. The equations for both resonance fields of the two split components are the following [71]:

$$H_r(1) = \frac{\omega_0}{\gamma} - \frac{6K_s}{RM} - H_a \qquad (6)$$

when the external field is parallel to the anisotropy axis, and

$$H_r(2) = \frac{\omega_0}{\gamma} + \frac{3K_s}{RM} + \frac{H_a}{2} \qquad (7)$$

when the field is perpendicular to the axis. In Eq. (6) and (7) $R$ stand for the radius of a nanoparticle. Subtracting Eqs. (6) and (7) and multiplying the result by 2/3, one gets

$$\frac{2}{3}[H_r(2) - H_r(1)] = \frac{6K_s}{RM} + H_a \qquad (8)$$

In our samples the spread of anisotropy axes is random so in fact only a single broad line is registered, but its decomposition on two components allow to find resonance fields in Eqs. (6) and (7). For GO-$Fe_3O_4$ sample the surface term (the first in the right-hand side in Eq. (8)) could be omitted as the bulk magnetocrystalline energy is supposed to be bigger than the surface energy. In that case the anisotropy field $H_a$ calculated from the line splitting (~180 Oe) in GO-$Fe_3O_4$ is equal to 120 Oe. This value is very close to the coercive field in this sample measured in isothermal magnetization. Actually, this is to be expected as for the coercive field $H_c$ the same equation as for $H_a$ is valid (see Eq. (5)). If the anisotropy field is the same in both polymer nanocomposites as in GO-$Fe_3O_4$, than Eq. (8) allows to calculate the surface anisotropy constant $K_s$. Substituting ~700 Oe for the line splitting, $R = 4.5$ nm, and $M = 90$ emu g$^{-1}$, the value of 0.15 erg cm$^{-2}$ for $K_s$ is obtained. The surface of magnetite nanoparticles can be strongly influenced by bonding to the polymer matrix.

A large shift of FMR lines from the reference field $H_{r0}$ in GO-$Fe_3O_4$ sample could be explained by agglomeration of magnetite nanoparticles in form of elongated assemblies along the lines of an external magnetic field. In contrast, in PTT-PTMO nanocomposites the polymer matrix prevents such rearrangements so the FMR shift is expected to be small or vanishing. Shape anisotropy field $H_S$ depends on the relative value of the three orthogonal demagnetization factors, $N_a$, $N_b$, and $N_c$, where $N_a + N_b + N_c = 4\pi$. For elongated particles or assemblies ($N_a = N_b$), $H_S = (N_a - N_c) \cdot M_S = \Delta N \cdot M_S$, where $M_S$ is the saturation magnetization. For elongated needles ($N_a = N_b = 2\pi$, $N_c = 0$), $H_S = 2\pi M_S$. The shape anisotropy field is adding as a vector to the external magnetic field. If elongated aggregate of nanoparticles is situated parallel to the external field directions, the following well known equations could be use:

$$\frac{\omega_0}{\gamma} = H_r + \Delta N \cdot M_S \qquad (9)$$

According to this equation the observed resonance field $H_r$ decreases by a factor proportional to the saturation magnetization and the difference of the demagnetization factors. The observed increase of the shift for GO-Fe$_3$O$_4$ sample with decreasing temperature can be explained by increase of magnetization. The value of $\Delta N$ calculated from Eq. (9) is approximately 0.10 for GO-Fe$_3$O$_4$. Although this particular approach to orientation dependence of the resonance field in FMR experiment seems adequate for our sample, more general one was previously reported [72, 73, 74].

In **Figure 16** the temperature dependence of the linewidths of the component lines for the three investigated samples is displayed. In general, all linewidths increase with decreasing temperature. This is usually attributed to a gradual suppression of the averaging effect of thermal fluctuations with decreasing temperature [75, 76]. The linewidths of two components in GO-Fe$_3$O$_4$ show almost linear increase during cooling down this sample from RT and the largest linewidth is found for the perpendicular component (comp. 2 in **Figure 16**). Line components linewidths in both nanocomposites display more complex thermal behavior, but a general tendency of increasing with temperature decrease is easily recognized, especially below 150 K. The observed temperature dependence of components linewidths shows a remarkable similarity to the temperature dependence of internal fields producing shift and splitting of FMR lines (see **Figure 15**). This is illustrated in **Figure 17** where the dependence of the component linewidths as a function of the line shift from the reference field $H_{r0}$ in GO-Fe$_3$O$_4$ and on the line splitting in the PTT-PTMO/GO-Fe$_3$O$_4$ nanocomposites is presented.

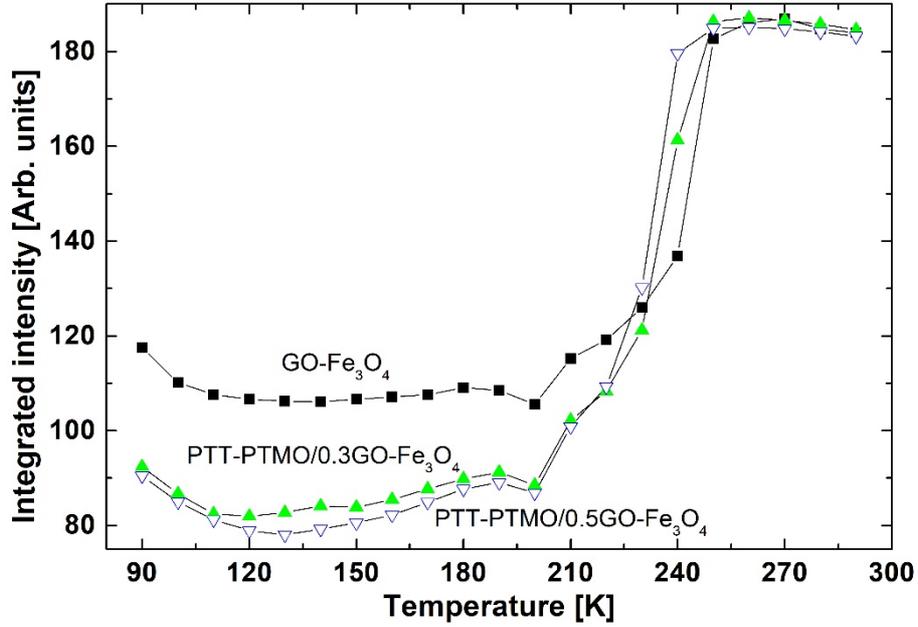

**Figure 18.** Temperature dependence of the normalized (at RT) integrated intensities for the GO-$Fe_3O_4$ and PTT-PTMO/GO-$Fe_3O_4$ nanocomposites. The solid lines are guides for the eyes.

The observed trend in linewidth broadening clearly indicate inhomogeneous broadening created by the spread of easy-axis directions in regard to appropriate internal field (demagnetization field in case of GO-$Fe_3O_4$ and surface anisotropy field in case of polymer nanocomposite samples) as a major contributor to the linewidth [20]. On the other hand, it could not be the only factor that determines the component linewidth because the observed linewidths are bigger than the appropriate internal fields. Other factors, like distribution of nanoparticle or cluster sizes should also have an effect on the observed component linewidth because size distribution influences directly the magnitude of the internal fields [75, 76].

In **Figure 18** the temperature dependence of the normalized (at RT) integrated FMR intensities $I_{FMR}$ (calculated as the sum of two components) for the PTT-PTMO/GO-$Fe_3O_4$ nanocomposites and GO-$Fe_3O_4$ is presented. $I_{FMR}(T)$ curves show a very interesting behavior. Initial cooling from RT causes a slight increase of $I_{FMR}$, but near 260 K a very rapid drop of FMR intensity is observed for GO-$Fe_3O_4$ and PTT-PTMO based nanocomposites. The magnitude of this decrease is slightly smaller for GO-$Fe_3O_4$ (40 % drop) than for both polymer nanocomposites (60 % drop). This might be explained by relatively more intense FMR spectrum of the latter samples at high temperatures in comparison to GO-Fe$_3$O$_4$ sample. Because the behavior of $I_{FMR}(T)$ is similar to $\chi_{ZFC}(T)$ it is possible to introduce the notion of FMR blocking temperature. It is surprisingly close to the blocking temperature determined from dc magnetization. FMR measurements should give a much higher value of the blocking temperature than determined from SQUID magnetization since the time windows $\tau$ for the two methods are very different: $\tau_{FMR} \sim 10^{-9}$ s, $\tau_{SQUID} \sim 10^2$ s [77]. But as was noticed by Ramos et al. [78] the notion of blocking in FMR has different meaning than in SQUID magnetization. That difference is the consequence of the fact, that in FMR the measurements are done in much stronger magnetic fields (~a few kOe) than in magnetization studies (only a few Oe). In consequence, if the effective anisotropy field becomes greater than $H_{r0}$ the nanoparticles whose easy axes are close to the direction of the applied field will no longer reach the resonance condition and a drop in FMR intensity will be registered. This idea might be supported by the observation that the components with the largest linewidth have the biggest contribution to the FMR intensity decrease below 260 K. Thus, the intensity decrease cannot be used as a feature that indicates the change from superparamagnetic to blocked regimes [79]. Because the magnitude of the internal anisotropy fields is similar in GO-Fe$_3$O$_4$ and PTT-PTMO/GO-Fe$_3$O$_4$ nanocomposites, FMR intensity drop is registered at a similar temperature ~260 K. On further cooling the $I_{FMR}$ curve levels down and below 120 K starts to slowly increase on temperature decrease. This increase is probably correlated with magnetization increase on temperature decrease observed in SQUID magnetization studies.

**Conclusions**

The polyester thermoplastic elastomeric nanocomposites with improved phase separated structure were obtained by *in situ* polymerization. Homogenous dispersion of GO-Fe$_3$O$_4$ nanoparticle hybrid in the PTT-PTMO matrix was confirmed by the TEM analysis. The average size of Fe$_3$O$_4$ nanocrystals distributed on GO sheets was ca. 9 nm. Due to presence of GO-Fe$_3$O$_4$ in copolymer matrix and better phase separation the improvement of tensile properties of PTT-PTMO/GO-Fe$_3$O$_4$ nanocomposites was observed. Superparamagnetic behavior for GO-Fe$_3$O$_4$ hybrid nanoparticles and nanocomposites with their content was observed at room temperature as well as hysteresis due to blocking of the magnetic dynamics at low temperature. A high value of 69.3 emu g$^{-1}$ for saturation magnetization in GO-Fe$_3$O$_4$ hybrid nanoparticles and 0.24 and 0.40 emu g$^{-1}$ for nanocomposites at loading of 0.3 and 0.5 wt% of GO-Fe$_3$O$_4$ in polyester thermoplastic elastomeric matrix, respectively, was registered.

Due to low toxicity and of GO-Fe$_3$O$_4$ hybrid nanoparticles [80, 81], the prepared elastomeric nanocomposites with their low loading could be developed for its potential application as a contrast agent in MRI, but this application requires further laboratory and in vivo investigations, especially with regard to the use of more biocompatible elastomer matrix. The used here PTT-PTMO thermoplastic elastomer as nanocomposite polymer matrix is more of an engineering material but it is possible to replace it by a more biocompatible elastomer, e.g. PTT-PEO copolymer [82] or PLA copolymer [81]. In surgical implants loaded with superparamagnetic GO-Fe$_3$O$_3$ hybrids acting as image contrast the sensitivity of MRI will be enhanced.

**Appendix A. Supplementary material**

**Acknowledgments**

Authors would like to thank DuPont Tate & Lyle Bio Products company for providing bio -1,3 propanediol for experimental use. Dris Ihiawakrim from IPCMS is acknowledged for performing TEM microscopy. Dr Matej Micusik from Polymer Institute of Slovak Academy of Sciences is acknowledged for XPS analysis. Zdenko Spitalsky gratefully acknowledge to VEGA 2/0093/16 for financial support.

**Conflict of Interest**

The authors declare no conflict of interest.